\documentclass{article}
\usepackage{geometry}
\usepackage{graphicx}
\usepackage{amsmath,accents}
\usepackage{amssymb}

\usepackage{hyperref}
\usepackage{xcolor}
\hypersetup{
    colorlinks,
    linkcolor={blue!80!black},
    citecolor={red!60!black},
    urlcolor={blue!80!black}
}
\usepackage{cite}
\usepackage{authblk}
\usepackage[skins]{tcolorbox}
\usepackage{parskip}
\usepackage{caption}
\def \Mp {M_{\rm P}}

\newcommand{\beq}{\begin{equation}}
\newcommand{\eeq}{\end{equation}}
\newcommand{\bea}{\begin{eqnarray}}
\newcommand{\eea}{\end{eqnarray}}

\newcommand{\aend}{a_{\rm end}}
\newcommand{\rhoend}{\rho_{\rm end}}

\title{\textbf{Phenomenological Constraints on Higgs reheating}}
\author[1]{Yann Cado}
\author[2]{Mathieu Gross}
\author[2]{Yann Mambrini}
\author[3]{Keith Olive}
\affil[1]{\small Laboratoire de Physique Th\'eorique et Hautes Energies (LPTHE), Sorbonne Universit\'e et CNRS UMR 7589, 4 place Jussieu, 75252 Paris CEDEX 05, France}
\affil[2]{Universit\'e  Paris-Saclay, CNRS/IN2P3, IJCLab, 91405 Orsay, France}
\affil[3]{William I.~Fine Theoretical Physics Institute, School of Physics and Astronomy, University of Minnesota, Minneapolis, MN 55455, USA}
\date{\today}

\begin{document}

\maketitle

\vspace{1cm}
\begin{abstract}
In many models of inflation, reheating is realized through a coupling between the inflaton and the Higgs boson. Often, the mass of the inflaton is of order $10^{13}$~GeV determined by the amplitude of the scalar fluctuation spectrum. However, in models where the inflaton potential is of the form $V \sim \phi^k$ about its minimum, the inflaton is massless for $k\ge 4$ unless a bare mass term, $\frac12 m_\phi^2 \phi^2$, is present. In this case, the inflaton mass may be of order the electroweak scale and may be subject to existing collider constraints. In particular, 
we investigate the constraints on the inflaton mass and reheating temperature $T_{\rm rh}$ arising from the decay of $\phi$ into  $\mathcal{H}$ through an interaction term $\mu \phi |\mathcal{H}|^2$.
  We perform a renormalization group analysis to determine the relative values of $\mu$ and $m_\phi$ such that the Higgs potential remains stable (and perturbative) at high energy.  
 Taking into account the running of the Higgs quartic self-coupling  and the experimental constraints from the LHC via the \texttt{HiggsTools} public code, we find that $3.4 \times 10^6 $ GeV $\lesssim T_{\rm rh}\lesssim   3.9 \times 10^{12} $ GeV with a corresponding constraint on the inflaton bare mass $260~{\rm GeV} \lesssim m_\phi \lesssim 3.8 \times 10^{10}~{\rm GeV}$. The dependencies between $T_{\rm rh}$ and the inflaton bare mass $m_\phi$ as well as between $\mu$ and $m_\phi$ are provided.

\end{abstract}

\vskip 3in

\begin{flushleft}
FTPI--MINN--25/08 \\
UMN--TH--4502/25
\end{flushleft}

\newpage
\tableofcontents

\section{Introduction}

The inflationary paradigm successfully addresses several shortcomings of the standard Big Bang cosmology, such as the horizon and flatness problems, by positing a period of accelerated expansion in the early Universe~\cite{Olive:1989nu,Linde:1990flp,Martin:2013tda,Martin:2013nzq,Martin:2015dha,Ellis:2023wic}. However, to be compatible with Big Bang Nucleosynthesis (BBN) (see e.g. \cite{Fields:2019pfx,Yeh:2022heq}), any inflation model must account for sufficient reheating after the period of exponential expansion ends.
Perturbative reheating is typically described by the decay of the inflaton field into daughter particles, e.g. scalars, fermions or gauge fields.
In many models of inflation, after inflation ends, the inflaton field begins to oscillate about the minimum of its potential. If the potential is approximately quadratic about the minimum, 
the coherent oscillations of the inflaton field are harmonic and behave like a condensate of inflaton particles with an average equation of state parameter $\langle w_\phi \rangle = p_\phi/\rho_\phi = 0$, typical of a  matter-dominated era. With an appropriate coupling to other fields, the inflaton can decay, and produce a thermal bath required for BBN \cite{Dolgov:1982th,Abbott:1982hn,Nanopoulos:1983up}.

Once specified, the scalar potential for the inflaton allows one to compute several cosmological observables associated with the cosmic microwave background (CMB) 
anisotropy spectrum. These place strong constraints on the shape and magnitude of the potential. For example, the Starobinsky potential \cite{Starobinsky:1980te} (derived 
originally from an $R+R^2$ modification of gravity) requires a very large mass
of the order of $10^{13}$ GeV for the inflaton to be consistent with the overall 
normalization of the CMB background. This potential predicts a value for the tilt of 
the anisotropy spectrum, $n_s$ and the tensor-to-scalar ratio, $r$ in good agreement 
with observations of the CMB \cite{Planck:2018vyg,Planck:2018jri}.
Reheating in this model occurs primarily through the decay of the inflaton (which arises as the extra degree of freedom in the $R+R^2$ theory) to the Standard Model Higgs boson (see e.g. \cite{Ema:2024sit}). 
The details of the inflaton decay strongly influence reheating dynamics and leave imprints on observables such as the number of inflationary $e$-folds and the spectrum of primordial perturbations \cite{Liddle:2003as,Martin:2010kz,Ellis:2015pla,Ellis:2021kad}.
 Despite its importance, the reheating epoch remains one of the least constrained phases of early Universe cosmology.

The class of potentials known as T-models \cite{Kallosh:2013hoa,Kallosh:2013yoa}, gives predictions for the inflation observables which are very similar to those made by the Starobinsky model and in good agreement with CMB observables. The scalar potential driving inflation for these models can be written as 
\beq\label{eq:attractor}
V_{\rm inf}(\phi) \;=\; \lambda_\phi M_P^4 \left[ \sqrt{6} \tanh \left(\frac{\phi}{\sqrt{6 \alpha_T}M_P}\right) \right]^k\, ,
\eeq
where $M_P = 2.435 \times 10^{18}$~GeV is the reduced Planck mass
and $\alpha_T$ is a model-dependent parameter
related to the curvature of the field-space manifold in supergravity models \cite{Ellis:2013nxa}. 
For $k \ge 4$, there is no direct (bare) contribution to the inflaton mass from Eq.~(\ref{eq:attractor}). However a bare mass term $\frac12 m_\phi^2 \phi^2$
can always be included along with the inflationary potential ~\cite{Clery:2024dlk}.
Notably, the bare mass of the inflaton field can not be excluded as it eventually arises from radiative corrections or some symmetry breaking that could affect the reheating and the low-energy behavior of the inflaton field \cite{Clery:2024dlk}.
As long as $m_\phi^2 \ll \lambda_\phi M_P^2 \left(\frac{\phi_{\rm end}}{M_P}\right)^{k-2}$, where $\phi_{\rm end}$ is the inflaton field value when inflation ends,  this term preserves the predictions of model with respect to the CMB observables.
Thus, in this class of models the bare mass of the inflaton is no longer 
directly constrained by the physics of inflation. This gives us an interesting degree of freedom for the physics of reheating.

Indeed, reheating in these models is often studied through the phenomenology of the decay products \cite{garcia:2020eof,Garcia:2020wiy,Garcia:2023obw}.  As in the Starobinsky model, we are primarily interested in the Higgs field as the main decay channel.
The high scale dynamics of the field is governed by the inflationary potential $V_{\rm inf}(\phi)$ which can reproduce the correct cosmological observables. The bare mass term, $m_\phi$, 
affects the reheating process, if sufficiently long. However, the coupling of the inflaton to the Higgs sector which ensures successful reheating will also have an impact on the Higgs potential and Higgs physics, 
restricting the allowed coupling between the two scalars, and therefore, the reheating temperature. 
The motivation of this work is precisely to link the reheating temperature to collider constraints,
derived from the coupling of the Higgs to the inflaton. We obtain constraints on the reheating temperature imposed by LHC data and the stability of the Higgs potential.
In Ref.~\cite{Clery:2024dlk}, the reheating temperature was studied in four cases 
(inflaton decay into scalars, fermions, vectors, and scattering to scalars). In this work, 
we concentrate solely on the reheating through the decay to the Higgs boson. 

As already noted, to decouple the low energy mass of the inflaton from the CMB observables,  we must take $k \ge 4$. In these cases, given sufficient time, the condensate will tend to fragment to inflaton quanta \cite{Garcia:2023dyf,Garcia:2023eol}. Without the presence of a bare mass term, if fragmentation is complete and rapid, 
reheating does not occur as the (almost) massless quanta do not decay. 
However, as shown 
in \cite{Garcia:2023dyf}, fragmentation is never complete and the inflaton maintains a 
small effective mass (which continues to decrease as the remaining condensate 
evolves). In the case of inflaton decay to fermions however, the decrease in mass (and hence the decay rate) is rapid enough to prevent the decay unless the initial decay 
rate was large enough to induce the decay before fragmentation. This typically 
requires a non-perturbative coupling of the inflaton to the fermion decay products
\cite{Garcia:2023dyf}.  In 
contrast, for the decay to scalars, and thus to the Higgs boson, the decay of the inflaton occurs despite fragmentation with only a small effect on the resulting reheating temperature. 
Aiming for a minimal approach, we will consider in this work 
that the reheating occurs only through the decay $\phi \to hh$ generated by a coupling $ \frac{1}{2}\mu \phi h^2$ in the Lagrangian.
As previously mentioned, this coupling governs the inflaton decay and the reheating process at high energies, while also generating $\phi-h$ mixing below the electroweak scale, which can affect the relevant Higgs physics in collider experiments.

Furthermore, it is well known that when radiative corrections in the Standard Model (SM) effective potential are considered, the 
value of the Higgs quartic coupling, $\lambda_0(h)$, runs negative at high energy, mainly from the contribution of 
the top quark, at the renormalization scale of $\mathcal{Q}_{I}\sim10^{11}$~GeV ~\cite{Degrassi:2012ry}. This is far below the energy scale at which reheating begins. 
This instability usually requires an ultraviolet (UV) completion of the model which 
modifies the relation between the low-energy and high-energy SM 
parameters~\cite{Bezrukov:2014ipa}.  In particular, there must be a modification in the value of the Higgs self-coupling at the reheating scale.
Consequently, another aspect of our work is dedicated to analyzing the stability of the Higgs, and in particular to providing the parameter space which allows for the inflaton decay $\phi\to hh$, without jeopardizing the stability or the unitarity of the Higgs sector. 
Indeed, we show that the cubic interaction $\mu \phi h^2$ while generating the reheating can also  stabilize the Higgs potential, as was done in Refs.~\cite{Barbon:2015fla,Cado:2022evn} in a different context.

The paper is organised as follows. In Section~\ref{sec:reheating} we introduce the dynamics of reheating through the $\phi \to hh$ coupling before analyzing the induced correction for the Higgs quartic coupling running in Section~\ref{sec:stability}. We  then tackle the question of the mass mixing at low energy in Section~\ref{sec:mixing} before deriving the constraints on the reheating temperature from the stability of the Higgs potential while retaining consistent reheating in Section~\ref{sec:discussion}. Our conclusions are summarized in Section \ref{sec:summary}.

\section{Post-inflationary dynamics, the modified Higgs potential and the reheating temperature}
\label{sec:reheating}
In this section, we briefly review the reheating process needed for our analysis. For a more complete discussion, we redirect the reader to Refs.~\cite{garcia:2020eof,Garcia:2020wiy}.
In an expanding background, the inflaton $\phi$ obeys the 
equation of motion 

\begin{equation}
\label{Eq:eom_inflaton}
\ddot{\phi}+3H\dot{\phi} +\frac{dV_\phi}{d\phi} \approx 0\,,
\end{equation}
where $H$ is the Hubble expansion rate and 
\begin{equation}
    \label{eq:Vinf}
V_\phi =   V_{\rm inf}(\phi) + \frac12 m_\phi^2 \phi^2 \, ,
\end{equation}
where
$V_{\rm inf}(\phi)$  
is given by Eq.~(\ref{eq:attractor}). Our results are not particularly dependent on the specific inflationary model, though we have in mind the T-model in Eq.~(\ref{eq:attractor}).
In the following, we will restrict our attention to $\alpha_T = 1$.

At the onset of the reheating stage, it is sufficient to expand $V_{\rm inf}$ about the minimum at $\phi = 0$, retaining only the first term in the expansion
\begin{equation}
    V_{\rm inf}(\phi)\sim \lambda_{\phi} M_{P}^{4}\left(\frac{\phi}{M_{P}}\right)^{k},  \qquad \phi \ll M_P \, . 
    \label{def:reheating-potential-approx}
\end{equation}
The solution to Eq.~\eqref{Eq:eom_inflaton} can then be parametrized as follows
\begin{equation}
\phi(t) = \phi_0(t)\underbrace{\sum_{\nu} \mathcal{P}_{\nu}(t)e^{-i\nu \omega t}}_{\mathcal{P}}\ ,
\label{Eq:generalphi}
\end{equation}
where $\phi_0$ is the envelope of the field and the $\mathcal{P}_{\nu}$ are the Fourier coefficients of the pseudo-harmonic function $\mathcal{P}$. Since the time scale for oscillations  $\mathcal{O}(m_{\phi,{\rm tot}})$, with
\beq
m_{\phi,\rm tot}^2=\left.\frac{\partial^2 V_\phi}{\partial \phi^2}\right|_{\phi=\phi_0}
=k(k-1)\lambda_\phi M_P^2\left(\frac{\phi_0}{M_P}\right)^{k-2}+m_\phi^2=m^2_{\phi,\rm eff}+m_\phi^2\,,
\eeq
is fast compared to the expansion rate of the universe, $H$, we can decompose the equation of motion (\ref{Eq:eom_inflaton}) between the envelope (the amplitude of oscillations) and the pseudo-periodic part
\begin{align}
    \dot{\phi_0} &= -\frac{6H}{k+2}\phi_0, \label{eq:envelope}\\
    \dot{\mathcal{P}}^{2} &= \frac{2m_{\phi,{\rm tot}}^{2}}{k(k-1)}(1-\mathcal{P}_{}^{k}). \label{eq:harmonics}
\end{align}

For all fields we will adopt the convention stated above for the mass terms
\begin{equation}
    \boxed{m^{2}_{X,\rm{tot}} = m^{2}_{X} + m^{2}_{X,\rm{eff} }\hspace{1cm} X\in\left(\phi,h\right)}\,,
\end{equation}
where $m_{X}$ refers to the bare mass of the particle while the terms $m_{X,{\rm eff}}$ refer to the interaction contribution to the total mass.

Solving Eqs.~\eqref{eq:envelope} and \eqref{eq:harmonics} leads to a decaying envelope and an oscillating frequency of the form
\begin{align}
    \phi_0 &= \phi_{\rm end} \left(\frac{a_{\rm end}}{a}\right)^{\frac{6}{k+2}},
    \label{phi0}\\
    \omega &\simeq m_{\phi,{\rm eff}} \ \sqrt{ \frac{ \pi k }{ 2(k - 1) }}\frac{ \Gamma\left( \frac{1}{2} + \frac{1}{k} \right) }{ \Gamma\left( \frac{1}{k} \right) } \,,
\end{align}
where $a_{\rm end}$ is the value of the scale factor at the end of inflation defined when $\ddot{a} = 0$. 
Note that this system of equations boils down to a sinusoidal function for
$k=2$ but develops anharmonicities for higher powers of the potential.
For the specific case $k=4$ which will be our benchmark point in the following,
this system of equation is approximately solved by
\begin{equation}
    \phi(t) \simeq 
    \phi_{\text{\rm end}} \left( \frac{a_{\text{\rm end}}}{a(t)} \right)
\rm{sn} \left( \frac{m_{\phi,eff}}{\sqrt{6}} (t - t_{\text{\rm end}}), -1 \right)\,,
\label{jac}
\end{equation}
and ${\cal P} = {\rm sn}(x,y)$ is the Jacobi elliptic function in Eq.~(\ref{jac}).
From this decomposition, the inflaton can be seen as an infinite sum of modes with energies $E = \nu\omega$ having a decreasing amplitude $\phi_{0\nu} = \phi_{0} \mathcal{P}_{\nu}$. 
From the particle physics point of view it is important to take into account the contributions of all the modes during the decaying phase of the inflaton since they give rise to some enhancement (or even suppression) depending on the interaction considered. However the overall dynamics can be obtained by averaging the oscillations to get the equation of state \cite{garcia:2020eof,Garcia:2020wiy}
\begin{equation}
    \langle w_{\phi}\rangle = \frac{k-2}{k+2},
\end{equation}
which gives $\langle w_\phi\rangle=\frac13$ for $k=4$.
We will later drop the brackets for convenience.

Note that if the duration of the reheating process is sufficiently long, the bare mass term will become relevant for before reheating is complete. This happens if $a_{\rm rh}> a_m$, where the scale factor at reheating is defined by
\beq
\rho_\phi(a_{\rm rh}) = \rho_{\rm R}(a_{\rm rh}) \, , 
\eeq
when the radiation density becomes equal to the energy remaining in the inflaton condensate. 
The scale factor when the bare mass term dominates, $a_m$, is defined by
\beq
\frac12 m_{\phi}^{2}\phi^{2}(a_{m}) = \lambda_{\phi}M_{p}^{4-k}\phi^{k}(a_{m})\, .
\eeq
Using 
\beq
\rho_\phi = \rho_{\rm end} \left(\frac{a_{\rm end}}{a} \right)^\frac{6k}{k+2} \qquad \rho_{\rm R} \simeq \rho_{\rm rh} \left( \frac{a_{\rm rh}}{a} \right)^\frac{6}{k+2} \, ,
\eeq 
where $\rho_{\rm end}$ is the inflaton energy density at the end of inflation and $\rho_{\rm rh} = (g_* \pi^2/30) T_{\rm rh}^4$ is the radiation density at reheating with
$g_* = 106.75$ relativistic degrees of freedom at reheating temperature $T_{\rm rh}$. 
Thus we have
\beq
\left( \frac{a_{\rm rh}}{a_{\rm end}} \right) = \left( \frac{\rho_{\rm end}}{\rho_{\rm rh}} \right)^\frac{k+2}{6k} .
\eeq

Using the solution (\ref{phi0}), we can easily obtain \cite{Clery:2024dlk}
\begin{equation}
     \frac{a_m}{a_{\rm end}} = \left( \frac{2 \lambda_\phi^{\frac{2}{k}} M_P^{\frac{2(4 - k)}{k}} \rho_{\text{end}}^{\frac{k - 2}{k}}}{m_\phi^2} \right)^{\frac{k + 2}{6k - 12}}\,.
\end{equation}

The coupling of the inflaton to the Higgs boson will impact the running of the Higgs quartic coupling $\lambda$.
As noted earlier, for $k=2$ the inflaton cannot solve the instability problem since its mass scale is fixed to $m_{\phi} \sim 10^{13}$~GeV by the Planck constraints on the amplitude of scalar fluctuations. We would then have $m_{\phi} \gg \mathcal{Q}_{I}$, making it impossible (within the context of the Standard Model) to avoid a period of Higgs instability, since the loop correction to $\lambda$ is generated by the exchange of virtual inflatons. This point is discussed in more detail in Sec.~\ref{sec:stability}.

 For the remainder of the paper we now set $k=4$.
 In this case, the coupling $\lambda_\phi$ is set from the amplitude of CMB fluctuations and can be approximated
 by \cite{Garcia:2020wiy,Clery:2024dlk}
 \beq
\lambda_\phi = \frac{\pi^2 A_s }{2N_*^2} = 3.3 \times 10^{-12}\qquad (k=4)\,,
\label{lambdaAs}
 \eeq
where $A_s = 2.1 \times 10^{-9}$ is the amplitude of the CMB spectrum \cite{Planck:2018vyg}. For $k=4$, the value of $N_* \simeq 56$ is fixed independent of the reheating temperature and $\rhoend^\frac14  = 4.8 \times 10^{15} $~GeV. 
 
To describe the reheating process through the decay of the inflaton to Higgs bosons, 
 we consider the most general renormalizable potential invariant under the $\mathbb Z_2$ 
 symmetry, $\phi\to -\phi$, with a soft breaking cubic term proportional 
 to $\mu$:
\begin{equation}
    V_{\rm tot}  = \lambda_\phi\phi^{4} + \frac{1}{2}m_{\phi}^{2}\phi^{2}+\lambda_{\phi h}\phi^2 |\mathcal{H}|^2 - \mu \phi |\mathcal{H}|^2 + V(\mathcal{H}),
    \label{eq:Vtot}
\end{equation}
where $V(\mathcal{H})$ is the SM Higgs potential, and 
\begin{equation}
    \mathcal{H} = 
\frac{1}{\sqrt{2}} \begin{pmatrix}
  0 \\
  h  \\
\end{pmatrix} , \hspace{2cm} \langle h\rangle=v\,\simeq\,246\text{ GeV}\,,
\label{Eq:higgs}
\end{equation}
$h$ being the physical Higgs field.
For large Higgs field configurations, we can neglect the bare mass
term of the Higgs compared to its self-coupling term in $V(\mathcal{H})$.

The parameters $\lambda_{\phi h}$ and $\lambda_\phi$ are constrained by the slow-roll conditions during inflation to very small values $\lambda_{\phi h},\lambda_\phi\ll 1$.
Their smallness is radiatively stable, as can easily be deduced from their one-loop $\beta$ functions
\begin{subequations} \begin{eqnarray}
\beta_{\lambda_{\phi h}}&=&\dfrac{\lambda_{\phi h}}{16\pi^2}\left[12\lambda_0+8\lambda_{\phi h}+24\lambda_\phi-\left( \dfrac{9}{2}g_2^2+\dfrac{9}{10}g_1^2-6 y_t^2  \right)   \right]\;\theta\left(\log \frac{\mathcal{Q}}{m_\phi}\right)\,,
\label{eq:RGE1}\\
\beta_{\lambda_\phi}&=&\dfrac{1}{16\pi^2}\,(2\lambda_{\phi h}^2+72 \lambda_\phi^2)\;\theta\left(\log \frac{\mathcal{Q}}{m_\phi}\right)\,,
\label{eq:RGE2}
\end{eqnarray} 
\end{subequations}
where $\theta(x)$ is the Heaviside function, equal to 1 (0) for $x\ge 0$ ($x< 0$). 
In our analysis, the reheating process is predominantly sourced by the decay $\phi\to hh$ and not the scattering process $\phi\phi\to hh$. We can then reasonably set 
$\lambda_{\phi h}=0$ without loss of generality, as this choice is technically natural at one loop\footnote{See Ref.~\cite{Ema:2017ckf} for a scenario where this coupling 
was taken into account.}. Moreover, from the measurement of the amplitude of density perturbations, we have  $\lambda_\phi\simeq 3.3 \times 10^{-12}$, see Eq.~\eqref{lambdaAs}. From
Eq.~\eqref{eq:RGE2}, we see that this value is very mildly affected by radiative corrections.

The decay rate arising from the coupling $\mu \phi |\mathcal{H}|^2$ is
\begin{equation}
\Gamma (\phi\to hh) = \frac{\mu_{\rm eff}^{2}}{32\pi m_{\phi,\rm tot}}\,,
    \label{eq:decay-reheating}
\end{equation}
where the effective coupling $\mu_{\rm eff}$ for $k=4$ is approximated\footnote{We have checked that this approximation numerically reproduces the results from \cite{Clery:2024dlk}.}  as \cite{Garcia:2020wiy}
\begin{equation}
\label{def:mueff}
    \mu_{\rm eff}^2 \approx \frac92 c' \frac{\omega}{m_{\phi}}\max\left(1,\frac{2m_{h,\rm tot}}{\omega}\right) \mu^{2}\,,
\end{equation}
with $c' \approx 0.37$  and where
\begin{align}
  m_{h,\rm eff} &= \sqrt{\mu \phi_0} 
  \label{eq:mheff}
\end{align}
is the effective mass of the Higgs boson\footnote{Note that Eq.~\eqref{def:mueff} is approximate, as a more careful computation taking into account the contributions 
from the complete set of anharmonic modes can be done\cite{Garcia:2020wiy} but is outside the scope of this paper.}. When $\mathcal{R}^{1/2} \simeq \frac{2m_{h,\rm eff}}{\omega} \gg 1$ the second argument of the max function accounts for the enhanced decay rate resulting from decays into bosonic final states.

\begin{figure}[ht!]
    \centering
\includegraphics[height = 5 cm]{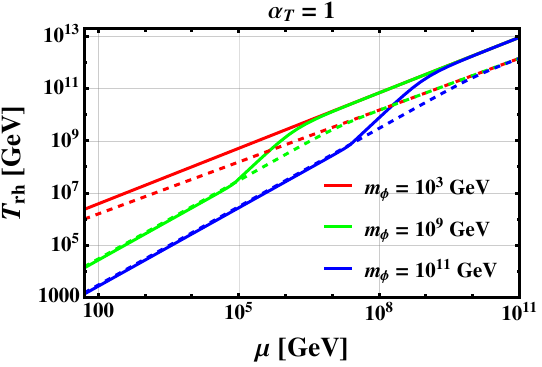}\hspace{5mm}
\includegraphics[height = 5 cm]{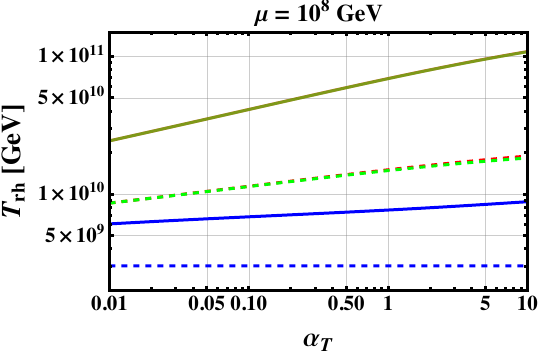}
    \caption{Left panel: The reheating temperature as a function of the parameter $\mu$ for different values of the inflaton bare mass $m_\phi$. Solid lines are obtained by solving numerically the Boltzmann equations \eqref{eq:systemRH}, while dashed lines are given by the analytical approximation \eqref{eq:Trh} without including the contribution of the effective mass. In both cases we reproduce the results from Ref.~\cite{Clery:2024dlk}.
    Right panel: The reheating temperature as a function of $\alpha_T$ for a fixed value of the coupling $\mu = 10^8$~GeV for the same choices of $m_{\phi}$. Note that in this panel the red and green lines overlap.
}
    \label{fig:reheating}
\end{figure}

It is important to clarify our procedure for calculating the 
reheating temperature. In Eqs.~\eqref{def:mueff} and \eqref{eq:mheff}, 
we need to take into 
account the running of the parameter $\mu$, computing it at the energy scale ${\mathcal Q}_{\rm rh}$ at which 
reheating takes place. 
In order to do so, we first fix a value of $\mu$ at the electroweak scale, 
and compute the reheating temperature $T_{\rm rh}^0$
{\it without} the running of $\mu$, to estimate 
the energy scale of reheating. 
Then, from the $\beta$ function of $\mu$, using Eqs.~\eqref{def:delta-lambda} and \eqref{eq:running-delta-lambda} below, 
we compute the value of $\mu(\cal Q)$ at the scale ${\cal Q}_{\rm rh}=T_{\rm rh}^0$, and reevaluate the new reheating temperature, $T_{\rm rh}^{\cal Q}$ implementing
the running value of $\mu({\cal Q}_{\rm rh}) $ in Eq.~(\ref{eq:decay-reheating}).
The difference between $T_{\rm rh}^0$ and $T_{\rm rh}^\mathcal{Q}$ depends on the
quartic couplings and $m_\phi$, and varies from $0.6\%$ to at most $15 \%$ in the range of the parameter space used in our study\footnote{The error value depends on the parameters $(\delta_\lambda,m_\phi)$ defined in Sec.~\ref{sec:stability}, and is small for small $\delta_\lambda$ and large $m_\phi$.  More precisely, the value $0.6\%$ corresponds to $\delta_\lambda = 0.05$, $m_\phi=10^{11}$~GeV while $15 \%$ corresponds to $\delta_\lambda = 0.25$, $m_\phi=10^4$~GeV.}.
Note that when we display values of $T_{\rm rh}$ in our figures, this refers to $T_{\rm rh}^{\cal Q}$,
and $\mu$ is the value at electroweak scale if no other scale is specified.
Note also that in order to reproduce the results in  Ref.~\cite{Clery:2024dlk} we plot $T_{\rm rh}^0$ in Fig.~\ref{fig:reheating} while in our final results, i.e. Figs.~\ref{fig:result1} and \ref{fig:result2}, we display the value of $T_{\rm rh}^\mathcal{Q}$.

To obtain $T_{\rm rh}^0$, and then $T_{\rm rh}^{\cal Q}$, we solve the coupled system of Boltzmann equations :
\begin{equation}
    \label{eq:systemRH}
    \begin{aligned}
&\dot{\rho}_{\phi}+3(1+w_{\phi})H\rho_{\phi} = -(1+w_{\phi})\Gamma_{\phi}\rho_{\phi},\\
    &\dot{\rho}_{R}+4H\rho_{ R} = (1+w_{\phi})\Gamma_{\phi}\rho_{\phi}\,,
 \end{aligned}
\end{equation}
where $\Gamma_{\phi}\simeq\Gamma (\phi\to hh)$ given by Eq.~(\ref{eq:decay-reheating})
is the leading channel.
The initial conditions at $a = \aend$ are $\rho_{R}(\aend) \approx 0$ 
and $\rho_\phi(\aend)=\rhoend$ as given above. Once these initial conditions are fixed, we solve the system numerically until reheating, defined by $\rho_{R}(a_{\rm{rh}}) = \rho_{\phi}(a_{\rm{rh}})$, which sets the reheating temperature $T_{\rm rh} =\left[\frac{30}{g_* \pi^2} \rho_{R}(a_{\rm{rh}})\right]^{ 1/4}$. After that point, the inflaton is exponentially damped and the Universe is radiation dominated. Note that exponential damping only occurs at $a > a_{\rm m}$ when the potential is dominated by the quadratic term.

Solving analytically the system of equations \eqref{eq:systemRH}, we obtain 
for $k=4$:
\begin{equation}
    \label{eq:Trh}
    T_{\rm rh} \simeq \left(\frac{30}{g_* \pi^2 } \right)^{\frac{1}{4}}\left[\frac{\mu_{\rm eff}^{2}\Mp}{144\pi^{2}\lambda^{\frac{1}{4}}} \right]^{\frac{1}{3}}\, .
\end{equation}
 Note that this relation holds only if reheating is completed before the mass becomes relevant (i.e., for $a < a_m$). We do not consider the possibility of reheating occurring at $ a > a_m$, as discussed in \cite{Clery:2024dlk}, due to our renormalization considerations, which ensure that such a scenario does not arise (see Sec.~\ref{sec:mixing}).
In the left panel of in Fig.~\ref{fig:reheating},   we show numerical solutions of the set of equations 
\eqref{eq:systemRH} for $T_{\rm rh}^0$ as a function of $\mu$, with (full lines) and without (dashed lines) taking into account  
the effective mass described by Eq.~(\ref{def:mueff}), 
for different values of the bare mass $m_\phi$.
 Fig.~\ref{fig:reheating} clearly exhibits a transition around $\mu \sim 10^{-3} m_{\phi}$ coming from the contribution of the effective mass captured by Eq.~\eqref{def:mueff} 
 which enhances the reheating temperature. 
 This corresponds to the value of $\mu$ such as $\mu \phi_0(a_{\rm rh}) \gtrsim m_\phi^2$
 before reheating is complete and to an enhancement 
 $\propto 
 \sqrt{\frac{\mu \phi_0}{m_\phi^2}}$ in $\Gamma_\phi$ in Eq.~(\ref{eq:decay-reheating}),  thus leading to an increase of the reheating temperature.
 This transition underlines the importance of 
 taking into account the effective coupling since it modifies $T_{\rm rh}$ by roughly one 
 order of magnitude.
 Next, we will see how this fundamental parameter for reheating,
 $\mu$, also affects (positively) the stability of the Higgs potential through  loop contributions 
 to  the Higgs quartic coupling $\lambda$. 
 
For completeness, we also show the influence of the inflationary parameter $\alpha_T$
in the right panel of Fig.~\ref{fig:reheating}.
We see that $\alpha_{T}$ influences the result by at most a factor of order one.
This subsequently justifies our choice of fixing $\alpha_{T} =1$
for the rest of our study.

\section{Constraints from vacuum stability}
\label{sec:stability}
\begin{figure}[ht!]
    \centering
      \includegraphics[height= 4.3cm]{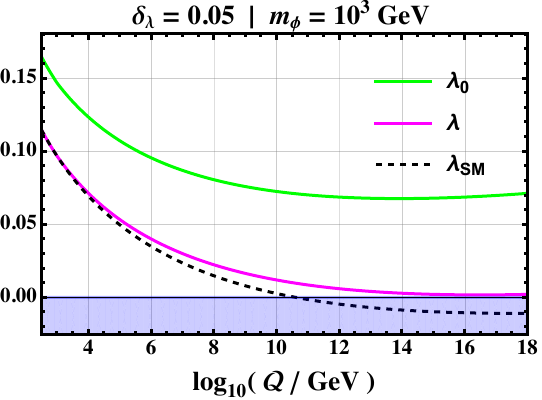} \hspace{1cm}
      \includegraphics[height= 4.3cm]{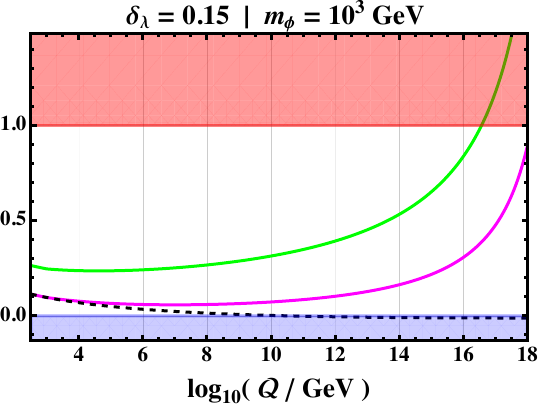}  \\\vspace{4mm}
      \includegraphics[height= 4.3cm]{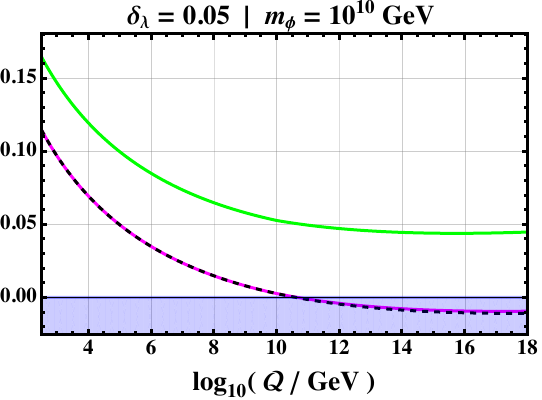} \hspace{1cm}
      \includegraphics[height= 4.3cm]{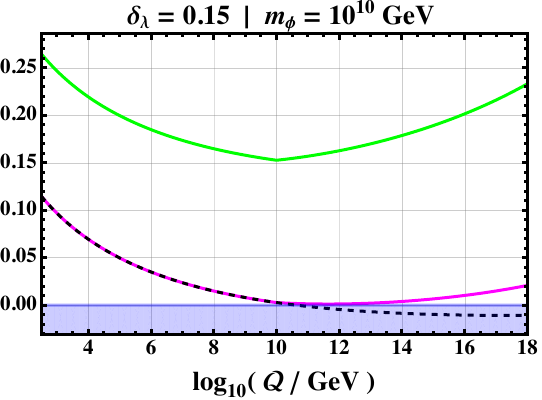}
    \caption{The Higgs self-coupling as a function of the  renormalization scale $\mathcal{Q}$ for four possible scenarios given by the values of $m_\phi$ and $\delta_\lambda$ evaluated at the scale $m_\phi$ i.e. $\delta_\lambda(m_\phi)$. The top left and the bottom right cases efficiently correct the SM running without overshooting $\lambda_0$. 
 Conversely, the top right case violates unitarity as $\lambda_0 >1$ when $\mathcal{Q}\gtrsim 10^{17}$~GeV and the bottom left case is unstable because $\lambda <0$ around $ \mathcal{Q}_I$. The colors of the excluded regions match the other plots. }
    \label{fig:running1}
\end{figure}

For the current values of the Higgs boson and top-quark masses, 
the SM potential becomes unstable when the Higgs field 
reaches $h\sim \mathcal Q_I \gtrsim 10^{11}$ GeV~\cite{Degrassi:2012ry}, as can be seen by the dashed black lines in Fig.~\ref{fig:running1}.
However, the coupling $ \mu \phi |\mathcal{H}|^2$, originally introduced to complete the reheating process,  will have an effect on the running of the Higgs self-coupling $\lambda$.

To begin with, one could expect a shift in the Higgs self-coupling at high-energy scale $\mathcal{Q}\equiv h>m_\phi$ due to the decoupling of the inflaton below ${\cal Q}\lesssim m_\phi$.
Indeed, at such a high energy scale, we can reasonably 
neglect the bare mass term in the Higgs potential and approximate it by 
\begin{equation}
V(h)\simeq \frac{1}{4}\lambda_0 h^4\,,
\end{equation}
where $\lambda_0$ is the (high energy) {\it bare} self-coupling that must be differentiated with the low energy SM self-coupling $\lambda$. Indeed, when $v<\mathcal Q<m_\phi$, the field $\phi$ decouples and is integrated out through the equation of motion for $\phi$, as
\begin{equation}
\frac{\partial V(\phi)}{\partial \phi}=0~~\Rightarrow ~~    \phi=\frac{\mu}{2m_\phi^2}h^2+\mathcal O(h^6)\,,
\label{eq:minC}
\end{equation}
which yields the potential
\begin{equation}
    V\simeq \frac{1}{2}m_\phi^2 \phi^2 -\frac{1}{2}\mu \phi h^2 + \frac{1}{4}\lambda_0 h^4 =  \frac{1}{4}\lambda h^4+\mathcal O(h^8)\,,
\label{eq:potC}
\end{equation}
where we neglected the quartic contribution to the inflationary potential. 
The high energy ($\lambda_0$) and low energy ($\lambda$)  self-couplings are then related by
\begin{equation}
    \lambda\equiv\lambda_0-\delta_\lambda\, ,
\label{def:lambda}
\end{equation}
with
\begin{equation}
\delta_\lambda\equiv\frac{\mu^2}{2m_\phi^2}\,.
\label{def:delta-lambda}
\end{equation}

The tree-level shift $\delta_\lambda$ increases the Higgs 
quartic coupling above $m_\phi$, $\lambda_0=\lambda+\delta \lambda$.
Nevertheless, the condition for stability requires 
\begin{eqnarray}
    \lambda(\mathcal{Q}) > 0, \text{ at all energy scales }\mathcal{Q}.
\end{eqnarray}
In Ref.~\cite{Elias-Miro:2012eoi}, it was argued that in models where there is a significant $\lambda_{\phi h}$ quartic coupling, threshold corrections may be sufficient to ensure the stability of the Higgs potential when $\lambda_{\phi h} > 0$, while stability is not achieved through threshold corrections with 
$\lambda_{\phi h} < 0$. It was later argued \cite{Barbon:2015fla} that in models with only a trilinear coupling ($\mu$), threshold effects are also insufficient for ensuring the stability of the potential. This is the case in the model considered here, as we are neglecting the quartic coupling $\lambda_{\phi h}$
(or assuming that it is small enough to be neglected).  Instead, it was further shown \cite{Barbon:2015fla} that at the loop level, the trilinear coupling may perserve the stability of the potential. 
The inflaton changes the renormalization group evolution of the Higgs quartic 
coupling above the threshold scale ${\cal Q}> m_\phi$.
As we will see, this effect is enough to stabilize the Higgs potential under certain conditions.

At low energies, the parameter $\lambda$ runs as the SM quartic coupling according to the SM $\beta$ function $\beta_{\lambda}^{\rm SM}$.
When $\mathcal Q>m_\phi$ however, the inflaton $\phi$ propagates and there is an extra contribution to the running of the parameter $\lambda$~\cite{Barbon:2015fla}.
That is,  modifications to the quartic coupling arise above the scale $m_{\phi}$ as the inflaton will start to affect the running introducing an extra contribution in the renormalization group equation~\cite{Barbon:2015fla} 
\begin{equation}
\beta_\lambda=\beta_\lambda^{\rm SM}+\frac{1}{2\pi^2}\delta_\lambda(3\lambda+\delta_\lambda)\,\theta\left(\log \frac{\mathcal{Q}}{m_\phi}\right). 
\label{eq:running-lambda}
\end{equation}
Here, we see explicitly how the running of the quartic coupling with the renormalization scale $\mathcal{Q}$ receives sizable {\it positive} contributions from the inflaton.
The parameter $\delta_\lambda$ also runs with the renormalization scale as~\cite{Cado:2022evn}
\begin{equation}
\beta_{\delta_{\lambda}}=\frac{1}{2\pi^2}\delta_\lambda(3\lambda+2\delta_\lambda)\;\theta\left(\log \frac{\mathcal{Q}}{m_\phi}\right),
    \label{eq:running-delta-lambda}
\end{equation}
which is why whenever a value of $\delta_{\lambda}$ is provided in this work, it is 
implied that this value is given at the energy scale $\mathcal{Q}=m_\phi$, i.e. $\delta_{\lambda}(m_\phi)$. The  value defined at low energy,
$\delta_{\lambda}(m_{\phi})$ is then used with both Eq.~\eqref{eq:running-lambda} and 
Eq.~\eqref{eq:running-delta-lambda} to test if the potential is  stable at the inflationary scale while taking into account the running of the standard model couplings\footnote{Note that we do not run $m_\phi$ in our set of equations for simplicity.}. We then deduce the reheating temperature (\ref{eq:Trh}), from the value of $\mu$, which is also fixed by $\delta \lambda$ in  Eq.~(\ref{def:delta-lambda}).

We show in Fig.~\ref{fig:running1} the evolution of the quartic couplings $\lambda$, $\lambda_0$ and $\lambda_{\rm SM}$ as function of the scale ${\cal Q}$, for different values of $\delta_\lambda$ and $m_\phi$.
Comparing the top panels ($\delta_\lambda=0.05$ versus $\delta_\lambda=0.15$)
for a low value of $m_\phi=10^3$ GeV, we see the strong influence 
of the inflaton on the stability of the Higgs. Even for low values such 
as $\delta_\lambda=0.05$, the issue of the Higgs instability can be solved for sufficiently low $m_\phi$. For $\delta_\lambda=0.15$, stability is easily achieved and the limit of the perturbativity bound ($\lambda_{0}\leq1$)\footnote{ Note that we have chosen this condition to ensure numerical stability in our computations. If one chooses a different value such as $\lambda_{0}\leq \sqrt{4\pi}$ it will open up the parameter space of Fig.~\ref{fig:running1} by raising the red shaded region.} 
 is reached for the bare quartic coupling,  $\lambda_0$,
before ${\cal Q}$ reaches the Planck scale. It is then clear that this will impact the reheating temperature, because upper bounds on $\delta_\lambda$ directly
imply an upper bound on $\mu$ through Eq.~(\ref{def:delta-lambda}),
which in turn implies an upper bound on the reheating temperature $T_{\rm rh}$
from Eq.~(\ref{eq:Trh}).


Increasing the value of $m_\phi$ has also an impact. It reduces the positive 
effect on $\lambda$, as one can see in the bottom panels of Figs.~\ref{fig:running1}. Indeed, larger $m_\phi$ means a larger range in ${\cal Q}$ for the SM running of  $\lambda$. The corrections due 
to the inflaton occur later (at higher energy ${\cal Q}$) so that the positive 
contribution to $\lambda$ is delayed, and furthermore is reduced at a given scale. 
Keeping in mind the same reasoning described above, for a given reheating temperature, on has an upper bound on $m_\phi$ to ensure the stability of the Higgs potential, even if the reheating is dominated by the quartic part of the inflaton potential.  Thus we see that for $m_\phi = 10^{10}$~GeV, the Higgs quartic coupling becomes negative for $\delta_\lambda = 0.05$. Increasing $\delta_\lambda$ to $\simeq 0.15$ restores the stability of the Higgs potential.

\begin{figure}[ht!]
    \centering
\includegraphics[height = 4.3 cm]{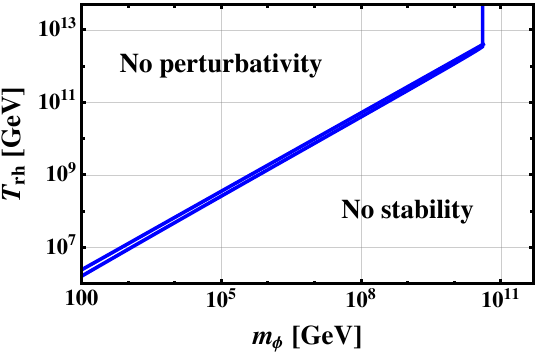}
    \caption{The reheating temperature as a function of the bare mass $m_\phi$. Lower mass values are constrained by collider phenomenology  and are not displayed here, see Sec.~\ref{sec:mixing} and Fig.~\ref{fig:result1}.
}
    \label{fig:result3}
\end{figure}

We show in Fig.~\ref{fig:result3} the two corresponding limits (stability and 
perturbativity) in the plane ($m_\phi$, $T_{\rm rh}$). We notice that the allowed 
region, between the two blue lines, is extremely narrow given the chosen ranges of the parameters. This translates to the fact 
that even a very small change in $\delta_\lambda$ drastically affects the stability of 
the Higgs, and can also easily lead to a non-perturbative coupling $\lambda_0$, 
as we just discussed. As a consequence, a very interesting point in our minimal framework, is the extreme predictivity of the model. The combination of these two
"formal" (in contrast to "experimental") bounds leads to a direct relation 
between $T_{\rm rh}$ and $m_\phi$.
Since the combination of vacuum stability and perturbativity fix $\delta_\lambda$ to a narrow range, for each value of $m_\phi$, $\mu$ is determined by $\delta_\lambda$. Furthermore, since $T_{\rm rh}$ is determined largely by $\mu$, we obtain the relation between $T_{\rm rh}$ and $m_\phi$.
This is one of the most important results of this work.

In summary, the extra contribution to the running of $\lambda$ in Eq.~(\ref{eq:running-lambda}) can solve the Higgs vacuum instability problem provided that:
\begin{itemize}
\item
The inflaton mass $m_\phi$ is smaller than the SM instability scale, $\mathcal Q_I\sim 10^{11}$~GeV.
\item
The value of $\delta_\lambda$ at the scale $\mathcal Q=m_\phi$, $\delta_\lambda(m_\phi)$, is large enough in order to significantly change the value of $\beta_\lambda^{\rm SM}$. 
\end{itemize}
One should also take care not to over-correct the running. There is an upper bound on the value of $\delta_\lambda(m_\phi)$ to guarantee that the theory remains in the perturbative regime up to the high scale.
Therefore we must require that $\lambda_0 < 1$ at all energy scales. This exclusion region is displayed in red in the corresponding plots. Note that then $\lambda < 1$ automatically holds.
Thus we obtain the approximate constraints $0.05 \lesssim \delta_\lambda\lesssim 0.2$ and $1 \text{ TeV} \lesssim m_\phi \lesssim 10^{10}$~GeV. Together, these imply that $300 \lesssim \mu / \text{GeV} \lesssim 6 \times 10^9$, in agreement with Ref.~\cite{Cado:2022evn}. We provide more precise numbers in Sec. \ref{sec:results}.

\section{Collider constraints on mass mixing}
\label{sec:mixing}
The presence of a $\mu \phi |{\cal H}|^2$ term also affects the low energy Higgs physics, through the $\phi-h$ mass mixing it generates.
Near the minimum of the potential, we can neglect the quartic part of $V(\phi)$.
The total potential for the Higgs and $\phi$ fields is then given by
\begin{equation}
V(\phi,\mathcal H)=\frac{1}{2}m_\phi^2\phi^2- \mu \phi |\mathcal H|^2 
-m_h^2|\mathcal H|^2+\lambda_0|\mathcal H|^4\,.
\end{equation}
%

The vacuum is defined as the solution to the minimum equations $\partial V/\partial \phi=\partial V/\partial h=0$, which provides $\langle h\rangle=v=246$~GeV and $\langle\phi\rangle=v_\phi$, with
\begin{equation}
m_h^2=\lambda v^2,\hspace{3cm} v_\phi=\sqrt{\dfrac{\delta_\lambda}{2}}\, \frac{v^2}{m_\phi},
\end{equation}
where the parameters $\lambda$ and $\delta_\lambda$ are defined in Eqs.~(\ref{def:lambda}) and (\ref{def:delta-lambda}), respectively.

The presence of the parameter $\delta_\lambda$ generates mixing between
$h$ and $\phi$, given by the squared mass matrix at the minimum 
\begin{equation}
\mathcal M^2=\begin{pmatrix} 2(\lambda +\delta_\lambda)v^2 & & -\sqrt{2\delta_\lambda}\,m_\phi v\\ -\sqrt{2\delta_\lambda}\,m_\phi v & & m_\phi^2 \,
\label{eq:M2}
\end{pmatrix}\,.
\end{equation}
Using the notation
\begin{equation}
    c_\alpha\equiv\cos\alpha, \hspace{1.5cm} s_\alpha\equiv\sin\alpha , \hspace{1.5cm}  t_\alpha\equiv\tan\alpha \,,
\end{equation}
this matrix is diagonalized by an orthogonal rotation with angle $\alpha$ as
\begin{equation}
\begin{pmatrix} c_\alpha & s_\alpha \\ -s_\alpha & c_\alpha \end{pmatrix} \mathcal M^2 \begin{pmatrix} c_\alpha & -s_\alpha \\ s_\alpha & c_\alpha \end{pmatrix}=\begin{pmatrix} m_{\tilde h}^2 & 0 \\ 0 & m_{\tilde\phi}^2 \end{pmatrix}\,,
\end{equation}
such that the mass eigenstates are
\begin{equation}
\tilde h=c_\alpha\, h+s_\alpha\, \phi,\hspace{3 cm} \tilde \phi=c_\alpha\, \phi-s_\alpha\, h\,,
\label{eq:physicalstates}
\end{equation}
and the mass eigenvalues are
\begin{equation}
\frac{m^2_{\tilde h,\, \tilde\phi}}{m_\phi^2}=\frac{1}{2}+\left(\lambda+\delta_\lambda  \right)\frac{v^2}{m_\phi^2}\mp
\sqrt{\frac{1}{4}-\left(\lambda-\delta_\lambda  \right)\frac{v^2}{m_\phi^2}+\left(\lambda+\delta_\lambda  \right)^2 \frac{v^4}{m_\phi^4}
}\,.
\label{eq:masas}
\end{equation}

In this way, the physical mass eigenstate $\tilde h$ is associated with the SM Higgs, with a mass $m_{\tilde h}=125.25$~GeV, while $\tilde\phi$ is a physical singlet, and both of them are coupled to the SM fields through the mixing angle $\alpha$. 
For $m_\phi\gg v$ the mixing angle is $s_\alpha\simeq \sqrt{2\delta_\lambda}( v/m_\phi)\ll 1$.

Hence, for $m_{\tilde \phi} \gtrsim 250$ GeV, we expect the existence of a 
scalar $\tilde \phi$ that decays mainly into the channel $\tilde\phi\to \tilde h \tilde h$ with a decay rate
\begin{equation}
\Gamma(\tilde\phi\to \tilde h\tilde h)=\frac{\kappa^2 \,m_\phi}{32\pi}\sqrt{1-\frac{4 m_{\tilde h}^2}{m_{\tilde \phi}^2}}\,,\hspace{1.5cm} \kappa=\sqrt{2\delta_\lambda}c_\alpha(1-3s_\alpha^2)+6 s_\alpha c_\alpha^2(\lambda+\delta_\lambda)\frac{v}{m_\phi}
\,.
\label{eq:Gamma-mixing}
\end{equation}
%
 The width of the resonance $\tilde \phi$ is typically around a few GeV in our region of interest.

At larger scales, like the one involved during the reheating process, 
the mixing vanishes, i.e. $\alpha \to 0$. Therefore $\kappa^2 m_\phi \to \mu^2/m_\phi$ and $\Gamma(\tilde\phi\to \tilde h\tilde h)$ reduces to $\Gamma (\phi\to hh)$ displayed in Eq.~\eqref{eq:decay-reheating}.
There are also subleading decay channels into SM particles, as $\tilde\phi\to X\bar X$ ($X\in \textrm{\rm SM}$), induced by the mixing with the Higgs boson, with very suppressed branching fractions 
\begin{equation}
\mathcal B(\tilde\phi\to X\bar X)=\mathcal B(\tilde h\to X\bar X) \times s^2_\alpha \;\frac{\Gamma_{\tilde h}}{\Gamma_{\tilde \phi}} \, \ll \, {\cal B}(\tilde \phi \to\tilde h\tilde h)
\,,
 \end{equation}
as $\Gamma_{\tilde h}\simeq 4 c^2_\alpha$ MeV in the SM, $\Gamma_{\tilde \phi}\simeq \Gamma(\tilde\phi\to \tilde h\tilde h)\simeq $ few GeV, so that $s^2_\alpha \Gamma_{\tilde h}/\Gamma_{\tilde \phi}\ll 1$.

In order to constrain the parameter space from collider data, we used the public code \texttt{HiggsTools} \cite{Bahl:2022igd}. \texttt{HiggsTools} has been developed to test beyond the Standard Model (BSM) physics with respect experimental results related to 125~GeV Higgs boson measurements at the LHC, with a particular focus on scenarios involving scalar particles.

More specifically, \texttt{HiggsTools} combines two complementary codes: \texttt{HiggsBounds} and \texttt{HiggsSignals}. The former takes Higgs-sector predictions as input and compares them with experimental searches for new particles at colliders to provide exclusion limits at the 95\% C.L. The latter confronts the predictions with measurements of the Higgs boson properties around $\sim 125$~GeV. In other words, \texttt{HiggsTools} takes our specific Higgs-sector model as input and returns two sets of limits, which may be independent. Our model must therefore be consistent with both \texttt{HiggsBounds} and \texttt{HiggsSignals} making our model compatible with the data from the Tevatron and the LHC.

\begin{figure}[ht!]
    \centering
    \includegraphics[height=4.5cm]{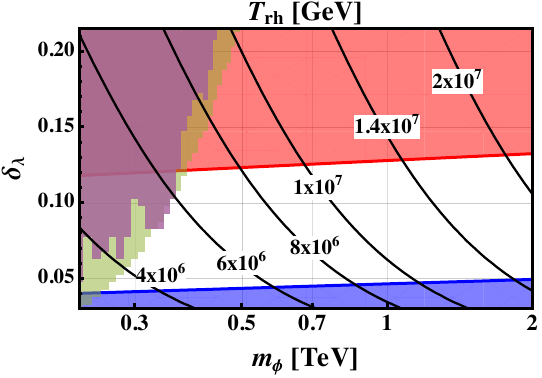}\hspace{5mm}
      \includegraphics[height=4.5cm]{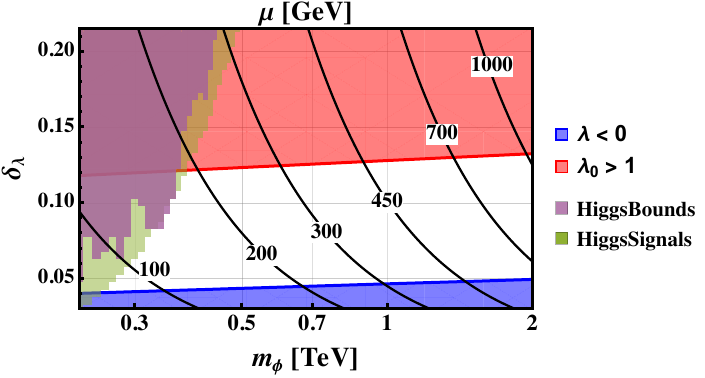}
    \caption{Contours of constant reheating temperature (left) and coupling $\mu$ (right) in solid black lines as a function of the inflaton bare mass focusing on low values of $m_\phi$. The shaded regions are excluded. We display the \texttt{HiggsTools} exclusion zone at 95\%C.L. and the perturbativity/stability conditions of the Higgs potential.}
    \label{fig:result1}
\end{figure}

To illustrate our results, we show in Fig.~\ref{fig:result1} the excluded regions at 95\%~C.L. from both \texttt{HiggsBounds} (in purple) and \texttt{HiggsSignals} (in green) in the plane
($m_\phi$, $\delta_\lambda$). We also plot  contours of constant $T_{\rm rh}$ ($\mu$)
in the left (right) panel.
We have also superimposed the perturbativity (red, top) and stability (blue, bottom) 
constraints. As expected, we see that the accelerator bounds restrict the low mass 
region for $m_\phi\lesssim 400$ GeV. This is when the mixing, $\sin \alpha$, begins 
to be sufficiently important to perturb the Higgs physics, within the experimental uncertainties.
The main  constraints come from the electroweak precision in the gauge boson
propagator, and the suppression of the Higgs boson decay due to the presence of the mixing angle. For more details, see Ref.~\cite{Cado:2022evn}. Note that lower reheating temperatures are reached only for low values of $\mu$
and hence low values of $\delta_\lambda$ well in the region where stability can not be affected by the coupling of the Higgs boson to the inflaton. 

To be more precise, we found that the mixing bounds obtained by present ATLAS and CMS data concerning the SM Higgs signal strengths, implies $m_\phi\gtrsim 220 \ (460)$~GeV for $\delta_\lambda=0.05\ (0.2)$. 
Our study involved a complete numerical analysis, using the codes \texttt{HiggsBounds} and \texttt{HiggsSignals}, but the overall verdict for the allowed parameter space $(m_\phi,\delta_\lambda)$ agrees well with the analytical estimate obtained in Ref.~\cite{Cado:2022evn}.
If we combined the LHC bounds with the limits from the Higgs self-coupling running obtained in the previous section, we can constrain directly the reheating temperature $T_{\rm rh}$, through $\mu$ in Eq.~(\ref{def:delta-lambda}).
This yields the minimum bounds for the reheating temperature as

\begin{center}\begin{tcolorbox}[enhanced,width=9cm,center upper,fontupper=\small,drop shadow southwest] 
$T_{\rm rh}\; \gtrsim \;3.4 \times 10^6 $ GeV $\quad $ for $m_\phi \simeq 260$ GeV \\\vspace{2mm}
$T_{\rm rh} \;\gtrsim\; 6.9 \times 10^6 $ GeV $\quad $ for $m_\phi \simeq 390$ GeV 
\end{tcolorbox}\end{center}
depending on the value of $\delta_\lambda$. The first lower bound corresponds to the minimum $m_\phi$ necessary
to ensure a stability while still not being excluded by LHC data, whereas the second lower bound respects the Higgs perturbativity and LHC constraints.
Note that 260 GeV can be considered as an {\it absolute} lower bound in the case of a quartic potential with a bare mass term for the inflaton.

\section{Results and discussion}
\label{sec:results}
\label{sec:discussion}

\begin{figure}[ht!]
    \centering
    \includegraphics[height=4.5cm]{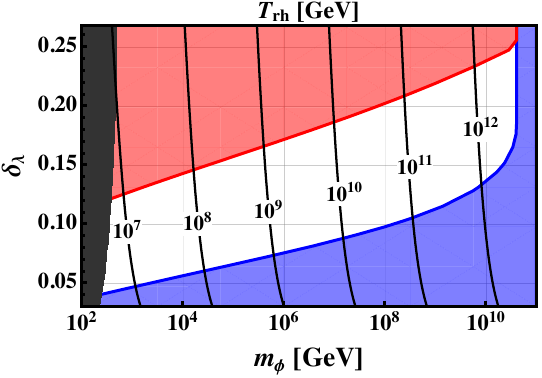}\hspace{5mm}
        \includegraphics[height=4.5cm]{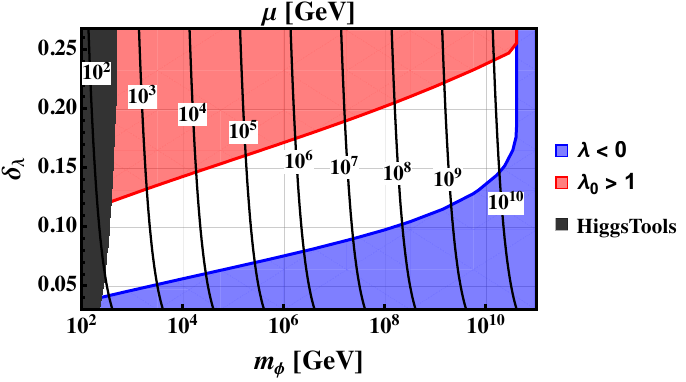}
    \caption{Contours of constant reheating temperature (left) and coupling $\mu$ (right) as a function of the inflaton bare mass for an extended range in $m_\phi$. The combined exclusion region from the collider constraints using \texttt{HiggsTools} are shown as the dark shading on left side of each panel. The two color regions displaying the perturbativity/stability of the Higgs potential are excluded.
    }
    \label{fig:result2}
\end{figure}

We are now in a position to combine our results from the two previous sections over the 
($m_\phi$, $\delta_\lambda$) parameter space. An extended range of parameters in this plane is shown in Fig.~\ref{fig:result2}.
Vacuum stability and perturbativity constrain $\delta_{\lambda}$ to a relatively small range.
The Higgs-inflaton mixing and the resulting collider bounds set a lower bound to the bare mass of the inflaton $m_\phi$, as already seen in Fig.~\ref{fig:result1}. This in turn, implies a {\it lower} bound on 
$\mu \gtrsim$ 74 GeV, at the TeV scale, due to the constraints given on $\delta_\lambda$ (\ref{def:delta-lambda}) from the Higgs {\it stability}, see 
right panel of Fig.~\ref{fig:result1}. The lower bound on $\mu$ can then be converted into
a lower bound on $T_{\rm rh}$ through Eq.~(\ref{eq:Trh}).
As a consequence, unless other new physics is introduced to stabilize the electroweak vacuum, the low reheating temperature regime, $T_{\rm rh} \lesssim 3.4 \times 10^6$ GeV is {\it forbidden} in this scenario. This is also one of the most important results of our work. A lower bound on $T_{\rm rh}$, above the PeV scale, is unavoidable if 
one wants to combine LHC constraints, with the stability of the Higgs potential at high energy.

In addition, we must also consider the perturbativity limit. For $\delta_\lambda$
respecting this limit ($\lesssim 0.25$ at the GUT scale,  see Fig.~\ref{fig:result2})  and for large $m_\phi$,   
the contribution of the inflaton to the quartic coupling to the Higgs, $\lambda$, is 
too weak to counterbalance the negative SM contributions before ${\cal Q}_I$, as one can see on the bottom-left panel of Fig.~\ref{fig:running1}. 
This, in turn, allows one to derive an {\it upper} bound on $\mu \lesssim 2.7\times 10^{10}$ GeV, which translates into an {\it upper}
bound on $T_{\rm rh} \lesssim 3.9 \times 10^{12}$ GeV, through Eq.~(\ref{eq:Trh}).
In summary, the combination of LHC, stability and perturbativity constraints 
places the following limits on the reheating temperature :
\begin{center}\begin{tcolorbox}[enhanced,width=9cm,center upper,fontupper=\small,drop shadow southwest] 
$3.4 \times 10^6 $ GeV $\;\lesssim \;T_{\rm rh}\;\lesssim \;  3.9 \times 10^{12} $ GeV\,,
\end{tcolorbox}\end{center}
which corresponds to a range for the inflaton mass
\begin{center}\begin{tcolorbox}[enhanced,width=9cm,center upper,fontupper=\small,drop shadow southwest] 
$260 $ GeV $\;\lesssim \;m_\phi\;\lesssim \;  3.8 \times 10^{10} $ GeV\,,
\end{tcolorbox}\end{center}
and for $\mu$ 
\begin{center}\begin{tcolorbox}[enhanced,width=9cm,center upper,fontupper=\small,drop shadow southwest] 
$74 $ GeV $\;\lesssim \; \mu \;\lesssim \;  2.7 \times 10^{10} $ GeV\,.
\end{tcolorbox}\end{center}

It is important to note that one cannot pick any arbitrary values of $m_{\phi}$ and $\mu$ within the intervals presented above and expect Higgs vacuum stability since the ratio of their squares should lie approximately between $0.1$ and $0.4$, coming from the limits on $\delta_\lambda$. 
The low mass limit coming from collider experiments should be seen as hard limit while the stability and perturbativity limit should be seen as softer ones: Stability can be achieved with other BSM physics entering below the scale ${\cal Q}_I$, and the perturbativity limit can perhaps be relaxed as we have taken the aggressive limit $\lambda \le 1$. 
The charm of our scenario is its minimalistic approach.
Nevertheless, the interplay between the inflaton and the Higgs boson tends {\it generically}
to point toward high values of the reheating temperature allowing processes that usually happen during radiation domination such as dark matter freeze out or baryogenesis for example.

\section{Conclusion}
\label{sec:summary}

Reheating after inflation is a necessary stage to achieve a successful transition to the standard radiation-dominated era. This process can occur through various mechanisms, such as particle physics interactions via specific couplings \cite{garcia:2020eof,Garcia:2020wiy} or through the evaporation of primordial black holes \cite{RiajulHaque:2023cqe}. In many approaches, the inflaton is assumed to couple to unspecified daughter particles, reflecting our limited knowledge of particle content at the inflationary energy scale $\rho_{\rm end}^{1/4} \sim 10^{15}$~GeV. This large separation from collider-accessible scales typically restricts observational probes of reheating to gravitational wave production, mostly in the high-frequency regime \cite{Kanemura:2023pnv,Bernal:2023wus,Choi:2024ilx,Garcia:2024zir,Xu:2024fjl,Bernal:2024jim,Gross:2024wkl,Xu:2024cey,Bernal:2025lxp,Xu:2025wjq}.

In inflationary models where the potential is quartic to first order in its expansion about its minimum, the inflaton is nominally massless, but the Lagrangian may naturally include a bare mass term, thus affecting the reheating process \cite{Clery:2024dlk}. In this context, the inflaton could become relevant at lower energy scales, making it more accessible to experimental constraints.

In this work, we investigated a minimal scenario in which the inflaton decays exclusively into the Higgs field while having a bare mass term. This setup is theoretically appealing as it provides a viable decay channel for the inflaton, preventing the formation of an unwanted relic abundance. Alternative scenarios, such as inflaton decay into fermions, can achieve similar results but is expected to only constrain the bare mass to be above the collider scale. Furthermore, this model has been shown to successfully account for baryogenesis \cite{Cado:2022evn} and address the metastability of the Higgs potential \cite{Barbon:2015fla}.

We have demonstrated that, due to its stabilizing effect on the Higgs potential, this scenario imposes limits on the reheating temperature in the range $10^6~\mathrm{GeV} \lesssim T_{\rm rh} \lesssim 10^{12}~\mathrm{GeV}$, assuming a correction to the quartic
Higgs coupling $\lambda$, $\delta_\lambda \sim 0.1$ \cite{Barbon:2015fla}. It is important to note that future collider experiments probing increasingly higher energy scales could raise the lower bound on $T_{\rm rh}$ in this model, thereby implying an extended period of radiation domination suitable for standard cosmological scenarios.

Our conclusions remain valid as long as the inflaton–Higgs decay is responsible for both reheating and the running of the Higgs quartic coupling. However, if additional couplings such as $\phi^2 h^2$ dominate the reheating process, the dynamics and constraints may differ. While such terms do affect the renormalisation group evolution of the Higgs coupling, they have been shown to only raise the instability scale \cite{Elias-Miro:2012eoi}.

\section*{Acknowledgements} \label{sec:acknowledgements}
We would like to thank Marcos Garcia for helpful discussions. 
This project has received support from the European Union's Horizon 2020 research and innovation program under the Marie Sklodowska-Curie grant agreement No 860881-HIDDeN.
Y.C. acknowledges funding support from the Initiative Physique des Infinis (IPI), a research training program of the Idex SUPER at Sorbonne Universit\'{e}. 
The work of K.A.O.~was supported in part by DOE grant DE-SC0011842  at the University of
Minnesota.
Y.M. acknowledges support by Institut Pascal at Université Paris-Saclay during the Paris-Saclay Astroparticle Symposium 2024, with the support of the P2IO Laboratory of Excellence (program “Investissements d’avenir” ANR-11-IDEX-0003-01 Paris-Saclay and ANR- 10-LABX-0038), the P2I axis of the Graduate School of Physics of Université Paris-Saclay, as well as the CNRS IRP UCMN.

\bibliographystyle{JHEP}
\bibliography{references}

\end{document}